\documentclass[preprint,aps,showpacs,preprintnumbers,amsmath,amssymb,nofootinbib]
{revtex4}
\usepackage{epsfig}
\begin{document}

\begin{flushright}
\end{flushright}


\newcommand{\be}{\begin{equation}}
\newcommand{\ee}{\end{equation}}
\newcommand{\bea}{\begin{eqnarray}}
\newcommand{\eea}{\end{eqnarray}}
\newcommand{\nn}{\nonumber}

\title{\large Neutrino mixing matrices with relatively large $\theta_{13}$ and with texture
one-zero}
\author{K. N. Deepthi, Srinu Gollu, R. Mohanta }
\affiliation{
School of Physics, University of Hyderabad, Hyderabad - 500 046, India }

\begin{abstract}
The recent T2K, MINOS and Double Chooz oscillation data hint a relatively large $\theta_{13}$, which can be
accommodated by some general modification of the Tribimaximal/Bimaximal/Democratic mixing
matrices. Using such  matrices we analyze several Majorana mass matrices with texture one-zero
and show whether they satisfy normal or inverted mass hierarchy and
 phenomenologically viable or not.
\end{abstract}

\maketitle

It is now well established by the recent neutrino oscillation
experiments \cite{ref1}
that neutrinos do have a tiny but finite nonzero mass.
Since neutrinos are massive, there will
be flavor mixing in the charged current
interaction of the leptons and a leptonic mixing matrix will appear
analogous to the CKM mixing matrix for the quarks. Thus, the  three
flavor eigenstates of neutrinos ($\nu_e,~ \nu_\mu,~ \nu_\tau$)
are related to the corresponding
mass eigenstates ($\nu_1,~ \nu_2,~ \nu_3$) by the unitary transformation

\bea
\left( \begin{array}{c}
 \nu_e       \\
\nu_\mu \\
\nu_\tau \\
\end{array}
\right ) \; =\; \left ( \begin{array}{ccc}
V_{e1}      & V_{e2}    & V_{e3} \\
V_{\mu 1}      & V_{\mu 2}    & V_{\mu 3} \\
V_{\tau 1}      & V_{\tau 2}    & V_{\tau 3} \\
\end{array}
\right ) \left ( \begin{array}{c}
\nu_1 \\
\nu_2 \\
\nu_3 \\
\end{array}
\right ) \; ,
\eea
where $V$ is the $3 \times 3 $ unitary
matrix known as PMNS matrix \cite{pmns}, which contains three mixing angles
and three CP violating phases (one Dirac type and two Majorana
type).  In the standard
parametrization \cite{ref20} the
mixing matrix is  described by  three mixing angles $\theta_{12}^{}$,
$\theta_{23}^{}$, $\theta_{13}^{}$ and three CP-violating phases
$\delta, \rho, \sigma$ as
\begin{eqnarray}
V = \left( \begin{array}{ccc} c^{}_{12} c^{}_{13} & s^{}_{12}
c^{}_{13} & s^{}_{13} e^{-i\delta} \\ -s^{}_{12} c^{}_{23} -
c^{}_{12} s^{}_{13} s^{}_{23} e^{i\delta} & c^{}_{12} c^{}_{23} -
s^{}_{12} s^{}_{13} s^{}_{23} e^{i\delta} & c^{}_{13} s^{}_{23} \\
s^{}_{12} s^{}_{23} - c^{}_{12} s^{}_{13} c^{}_{23} e^{i\delta} &
-c^{}_{12} s^{}_{23} - s^{}_{12} s^{}_{13} c^{}_{23} e^{i\delta} &
c^{}_{13} c^{}_{23} \end{array} \right) P^{}_\nu \;
,\label{mat}
\end{eqnarray}
where $c^{}_{ij}\equiv \cos \theta^{}_{ij}$, $s^{}_{ij} \equiv \sin
\theta^{}_{ij}$ and $P_\nu^{} \equiv \{ e^{i\rho}, e^{i\sigma}, 1\}$ is
a diagonal matrix with CP violating Majorana phases $\rho$ and $\sigma$.
The global analysis of the recent results of various neutrino oscillation experiments
\cite{ref21} suggest the neutrino masses and mixing parameters at $1 \sigma~(3 \sigma)$
level to be
\begin{eqnarray}
&&\Delta m^2_{21} = 7.59\pm 0.20(^{+0.61}_{-0.69})  \times 10^{-5}
\mbox{eV}^2\;,\nonumber\\
&&\Delta m_{31}^2 =
\left\{\begin{array}{ll}
+(2.46\pm0.12(\pm 0.37))\times 10^{-3}_{}~~\mbox{eV}^2 &
~~{\rm for~normal~hierarchy~(NH)} \\
-(2.36\pm 0.11(\pm0.37)) \times 10^{-3}_{}~~\mbox{eV}^2 &
~~{\rm for~inverted~hierarchy~(IH)}
\end{array}\right. ,\nonumber \\
&&\theta_{12} = 34.5\pm 1.0(^{+3.2}_{-2.8})^\circ\;,\;\;\theta_{23}
= 42.8^{+4.7}_{-2.9}(^{+10.7}_{-7.3})^\circ\;,\;\;\theta_{13} =
5.1^{+3.0}_{-3.3}(\le 12.0)^\circ\; ,
\end{eqnarray}
and as per the the latest ${\rm T2K} $  result \cite{ref22},
$\theta_{13}^{}$ at $90\% $ confidence level is found to be
\begin{eqnarray}
&  & 5.0^\circ \lesssim \theta^{}_{13} \lesssim 16.0^\circ ~~~~
({\rm ~NH~}) ,
\nonumber \\
&  & 5.8^\circ \lesssim \theta^{}_{13} \lesssim 17.8^\circ ~~~~
(~{\rm IH}~),
\end{eqnarray}
for a vanishing Dirac CP-violating phase $\delta $.
Moreover, the best-fit value of $\theta_{13}^{}$ is
found to be $\theta_{13}^{}\simeq 9.7^\circ$ for NH
and $\theta_{13}^{}\simeq 11.0^\circ$ for IH.
Soon after the T2K report, the MINOS Collaboration \cite{minos} has also released new data which
indicate $\theta_{13}\neq 0^\circ$ at the $1.5\sigma$ level.
Furthermore, evidence for  $\theta_{13}\neq 0^\circ$ at about  $3\sigma$ level has been
obtained in a global analysis \cite{valle, fogli}.
The first result from Double Chooz experiment \cite{de} reported at the LowNu conference as
\be
\sin^2 2 \theta_{13}=0.085 \pm 0.029~{\rm (stat.)} \pm 0.042 ~{\rm (syst.)}\;,
\ee
which also hints towards a non-zero $\theta_{13}$.
This issue has already been discussed in the literature
by several groups \cite{xing,
hezee,zheng,Manst,zhou,Araki,haba,meloni,sterli,ref23,ref24,ref25,hjhe}.
If this mixing angle is confirmed to be not very small by
new data from these two experiments and upcoming reactor neutrino
experiments, it will  provide an important constraint on theoretical model
building for neutrino mixing.

In the absence of any convincing flavor theory several approaches have
been proposed  to study the flavor problems
of massive neutrinos e.g., radiative mechanisms, texture zeros, flavor symmetries, seesaw
mechanisms, extra dimensions etc. The neutrino mass matrices with texture-zeros
are phenomenologically very useful as they allow the possibility
of calculating the neutrino mass matrix $M_\nu$  from which both the neutrino
mass spectrum and the flavor mixing pattern can be more or less predicted.
Theoretically, various neutrino mixing patterns have been proposed
using discrete flavor symmetries, e.g., $A_4$, $\mu -\tau$ symmetry etc.
Among those there are three well established patterns which are
of special interest: bimaximal mixing pattern (BM) \cite{bi},
tri-bimaximal mixing pattern
(TB) \cite{tri} and
democratic mixing pattern (DC) \cite{minzhu}, whose explicit forms are given below
\begin{eqnarray}
V_{\rm TB} &=& \left ( \begin{array}{ccc}
\sqrt{2\over 3}&\sqrt{1\over 3}&0\\
-\sqrt{{1\over 6}}&\sqrt{{1\over 3}}&\sqrt{1\over 2}\\
-\sqrt{{1\over 6}}&\sqrt{{1\over 3}}&-\sqrt{{1\over 2}}
\end{array}
\right )\; , \hspace{1cm}
V_{\rm BM}=\left(
\begin{array}{ccc}
\sqrt{1\over 2 } & \sqrt{1\over 2 } & 0 \\
-{1 \over 2 }& {1 \over 2 } & \sqrt{1\over 2 } \\
{1 \over 2 } &  -{ 1 \over 2 } &\sqrt{1\over 2 }
\end{array}\right) \;,\nn\\
V_{\rm DC} & = & \left ( \begin{array}{ccc}
\sqrt{\frac{1}{2}}&\sqrt{\frac{1}{2}}&0\\
\sqrt{\frac{1}{6}}&-\sqrt{\frac{1}{6}}&-\sqrt{\frac{2}{3}}\\
-\sqrt{\frac{1}{3}}&\sqrt{\frac{1}{3}}&-\sqrt{\frac{1}{3}}
\end{array}
\right )\;. \label{vtri}
\end{eqnarray}
Clearly, all these patterns suggest vanishing $\theta_{13}$
contradicting the recent observations of $\theta_{13}$ being
considerably large. Thus, to accommodate large $\theta_{13}$
 Refs. \cite{xing}  and \cite{hezee}
have considered  possible perturbations to the Democratic
neutrino mixing pattern and to
 the tri-bimaximal mixing pattern respectively.
 In Ref.  \cite{rf10}  it has been shown that if one assumes some general modification of the
neutrino BM/TBM/DC mixing pattern  then it is possible to get appropriate
neutrino mixing angles that may fit the T2K data.
The possible modification \cite{rf10} could be of the following forms
\begin{eqnarray}
&&1.~ V_{\rm PMNS}= V_{\alpha}^{} \cdot V_{ij}^{} \; , \label{p1}\\
&&2.~  V_{\rm PMNS}= V_{ij}^{} \cdot  V_{\alpha}^{} \; ,\label{p2} \\
&& 3.~V_{\rm PMNS}= V_{\alpha}^{} \cdot V_{ij}^{} \cdot V^{}_{kl} \; ,\label{p3} \\
&& 4. ~ V_{\rm PMNS} =  V_{ij}^{} \cdot V^{}_{kl} \cdot
V_{\alpha}^{} \; ,\label{p4}
\end{eqnarray}
where $\alpha = {\rm TB, BM~or ~ DC }$ and $(ij), (kl) =(12), (13),
(23)$ respectively. The perturbation mixing matrices
$V^{}_{ij}$ are given by
\begin{eqnarray}
V^{}_{12} &=& \left (\begin{array}{ccc}
\cos x & \sin x &0\\
-\sin x &\cos x &0\\
0&0&1
\end{array}
\right )\;, \;\;~~~~~~~~~~~~~ V^{}_{23} = \left (\begin{array}{ccc}
1&0&0\\
0&\cos y  &\sin y ~e^{i \delta}\\
0&-\sin y ~e^{-i \delta}& \cos y
\end{array}\right )\nn\\
V^{}_{13} & = & \left ( \begin{array}{ccc}
\cos z &0&\sin z ~ e^{i\delta} \\
0&1&0\\
-\sin z ~e^{-i \delta} &0& \cos z
\end{array}
\right )\;. \label{vb}
\end{eqnarray}

However, this modification is not the only way to obtain the large reactor angle $\theta_{13}$.
It has been shown in Ref. \cite{king} that a non-zero $\theta_{13}$ can arise at the
leading order  from type-1 see saw mechanism in the extension of TBM mixing matrix with partially
constrained sequential dominance. Such partially constrained sequential dominance can be
realized in the GUT models with a non-abelian discrete family symmetry, such as $A_4$,
spontaneously broken by flavons with particular vacuum alignment. Recently, it has also
been discussed in \cite{ketan} that it could be possible to achieve large $\theta_{13}$
through the deviation of from the exact TBM mixing, such as in a model with $S_4$ flavor
symmetry.

In the present work, we study the phenomenological implications of the above
mixing matrices with some texture one-zero structure in the neutrino mass matrices.
Such type of neutrino mass matrices with one/two texture zeros or one/two vanishing minors with $\mu-\tau$ symmetry
are discussed in Ref. \cite{rf11} where they have imposed  additional  constraint of TBM mixing.

It is well known that in the basis where the charged lepton mass matrix is diagonal, the complex symmetric mass matrix $M_{\nu}$ can be diagonalized by unitary matrix $V$ as
\begin{equation}
M_{\nu}=V M_{\nu}^{diag}V^{T}.\label{mass}
\end{equation}
where
\begin{center}
$M_{\nu}^{diag}$ = $\left(
\begin{array}{ccc}
m_{1} & 0 & 0 \\  0& m_{2} & 0 \\ 0& 0 & m_{3}
\end{array}
\right)$\;.
\end{center}
Thus, from (\ref{mass}) we have matrix elements of $M_{\nu}$ as
\begin{eqnarray}
({M_{\nu}})_{11} &=& m_{1}V_{11}^2+m_{2}V^{2}_{12}+m_{3}V^{2}_{13}\;,\nonumber \\
(M_{\nu})_{12} &=& m_{1}V_{11}V_{21}+m_{2}V_{12}V_{22}+m_{3}V_{13}V_{23}\;,\nonumber \\
(M_{\nu})_{13}& = & m_{1}V_{11}V_{31}+m_{2}V_{12}V_{32}+m_{3}V_{13}V_{33}\;,\nonumber \\
(M_{\nu})_{21}& = &m_{1}V_{11}V_{21}+m_{2}V_{12}V_{22}+m_{3}V_{13}V_{23}\;,\nonumber \\
(M_{\nu})_{22} &= &m_{1}V_{21}^2+m_{2}V^{2}_{22}+m_{3}V^{2}_{23}\;,\nn\\
(M_{\nu})_{23}& = & m_{1}V_{21}V_{31}+m_{2}V_{22}V_{32}+m_{3}V_{23}V_{33}\;,\nonumber \\
(M_{\nu})_{31} &= & m_{1}V_{11}V_{31}+m_{2}V_{12}V_{32}+m_{3}V_{13}V_{33}\;,\nonumber \\
(M_{\nu})_{32} &= & m_{1}V_{21}V_{31}+m_{2}V_{22}V_{32}+m_{3}V_{23}V_{33}\;,\nonumber \\
(M_{\nu})_{33} &= & m_{1}V_{31}^2+m_{2}V^{2}_{32}+m_{3}V^{2}_{33}\;.
\end{eqnarray}
The above mass matrix no longer respects  $\mu-\tau$ symmetry, however it would
satisfy  the $\mu-\tau$ symmetry property in the limit $V_{2i}=V_{3i}$.

Now we apply the six possible texture one-zero  condition to  $M_{\nu}$.  By doing so we can further filter which modifications to TB, BM, DC mixing patterns will give results comparable to experimental results.
Furthermore we choose the CP violating phases  $\delta$, $\rho$ and $\sigma$ to be
zero for simplicity in our analysis.

 We first consider the phenomenologically feasible perturbations to TB mixing pattern
i.e.,  $ V_{PMNS}= V_{TB}.V_{23}.V_{12}$, where $V_{TB}$ and $V_{23(12)}$ are given
in (\ref{vtri}) and (\ref{vb}). Comparing the resulting mixing matrix with its standard parametrization
form (\ref{mat}), we obtain the mixing angles as
\begin{eqnarray}
\nonumber
\tan\theta_{12}&=&\left|\frac{\cos x\cos y+\sqrt{2}\sin x}{\sqrt{2}\cos x-\cos y\sin x}\right|,\\
\nonumber
\sin\theta_{13}&=&\left|\frac{1}{\sqrt{3}}\sin y\right|,\\
\tan\theta_{23}&=&\left|\frac{\sqrt{3}\cos y+\sqrt{2}\sin
y}{-\sqrt{3}\cos y+\sqrt{2}\sin y}\right|,
\end{eqnarray}
By substituting the values of $ \theta_{23} $ and $ \theta_{12} $ we obtained the values of $x$
and $y$ as
\be
 x=\left (33.2^{+1.4}_{-3.2}\right )^\circ \;,~~~~y= -\left (88.2_{-3.1}^{+1.7} \right )^\circ \;. \label{xy}
 \ee
Using the above mixing matrix, the elements of neutrino mass matrix $ M_{\nu}$ can
be obtained from (\ref{mass}). Now we consider
the following texture one-zero pattern
 on the above $ M_{\nu} $ matrix elements, and see whether they are phenomenologically viable or not.

{\bf{Case 1: $(M_{\nu})_{11}=0$}}

By considering the (1,1) element of the neutrino mass matrix to be zero we obtain the condition that
\be
m_{2}(\sqrt{3}\cos x \cos y+\sqrt{6} \sin x)^2+m_{1}(\sqrt{6}\cos x -\sqrt{3} \cos y \sin x)^2
 + 3 m_{3} \sin^2y=0 \;.\label{eqm11}
 \ee

 From this equation it is not possible to infer any conclusion regarding the ratio of the neutrino
 masses as it involves masse terms of all three neutrinos. However, one can obtain the results
 by assuming the hierarchical nature of neutrino masses. For example for normal hierarchy where
 the neutrino masses follow the pattern $m_1 < m_2 << m_3$. Since the absolute scale of the neutrino
 mass is not precisely known, we assume the limit $m_1  \to 0$ for normal hierarchy. Similarly
 for inverted hierarchy which has the mass ordering as  $m_3< < m_1 < m_2$, we assume the limit
 $m_3 \to 0$.

With these assumptions, thus, we find for Normal Hierarchy i.e., \ $m_{1}\rightarrow 0$
\be
\dfrac{m_{2}}{m_{3}} =\left |\dfrac{\sin^2{y}}{\cos^2{x}\cos^2{y}+2\sin^2{x}+\sqrt{2}\cos{y}\sin{2x}}\right |\;,
\ee
and for  Inverted Hierarchy i.e., in the limit  $m_{3}\rightarrow 0$,
\be
\dfrac{m_{1}}{m_{2}}=\left |\dfrac{\cos^2 x \cos^2y+2\sin^2x+\sqrt{2}\cos y \sin2x}
{2\cos^2x-\sqrt{2}\sin 2x \cos y +\sin^2 x \cos^2 y }\right |\;.
\ee
By substituting the values of $x$ and $y$, as given in Eq. (\ref{xy}) we obtain the ratio of the masses as
and the variation of $m_2/m_3$ and $m_1/m_2$ with the perturbation parameter $x$ are shown
in Figure-1. From the figure, one can conclude that
\bea
&& \dfrac{m_{2}}{m_{3}}> 1~~~~~{(\rm for ~~NH)}\;,\nn\\
&& \dfrac{m_{1}}{m_{2}}< 1 ~~~~~~{(\rm for ~~IH)}.
\eea
\begin{figure}[htb]
\includegraphics[width=8cm,height=6cm, clip]{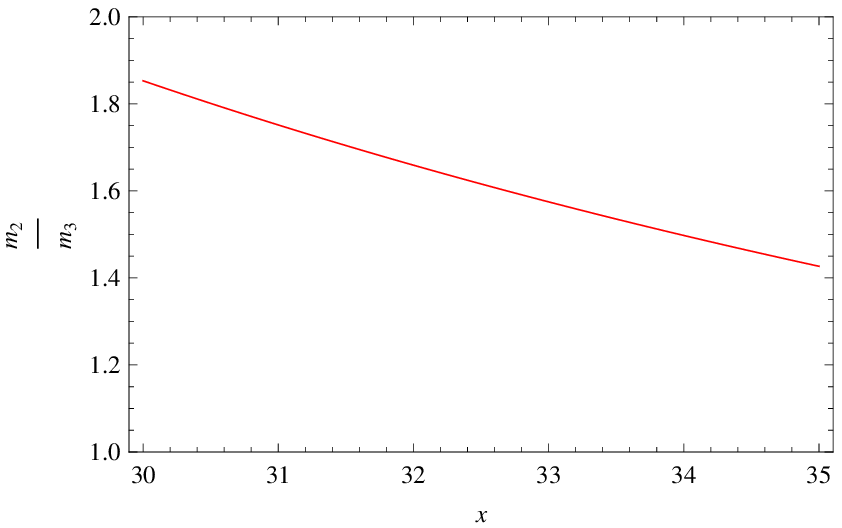}
\hspace{0.2 cm}
\includegraphics[width=8cm,height=6cm, clip]{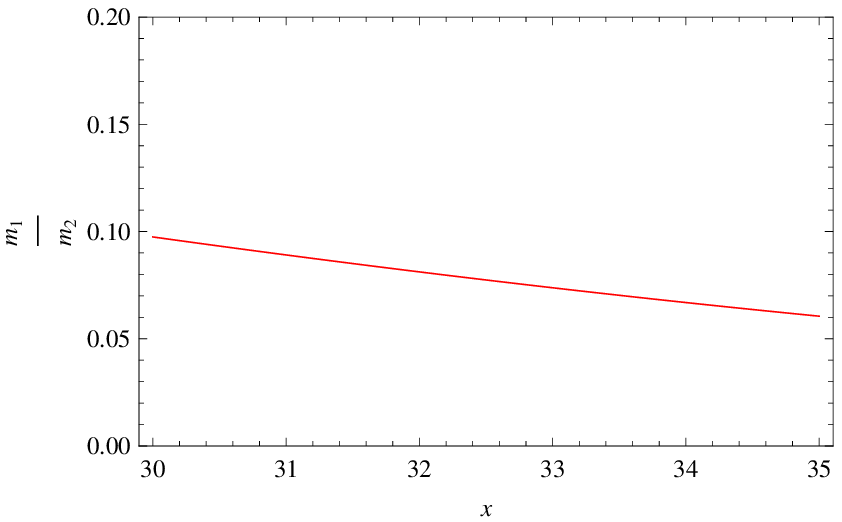}
\caption{Variation of the neutrino mass ratios $m_2/m_3$  with
the perturbation parameter $x$
(left panel) and $m_1/m_2$ on the right panel.}
\end{figure}

Thus, the above analysis shows that the neutrino mass matrix having texture one-zero with
vanishing $(M_\nu)_{11}$ will follow Inverted Mass Hierarchy pattern. Now since
we know that texture-zero type with $(M_{\nu})_{11}$ supports Inverted mass hierarchy
pattern, it is possible to determine the absolute mass scale of the lightest neutrino
i.e., $m_3$ from Eq. (\ref{eqm11}). Now writing Eq. (\ref{eqm11}) in the symbolic
form
\bea
a m_1  +c m_3=-b m_2\;,\label{eqm11a}
\eea
where $a$, $b$ and $c$ are the coefficients of $m_1$, $m_2$ and $m_3$ in (\ref{eqm11})
(e.g., $a= (\sqrt{6}\cos x -\sqrt{3} \cos y \sin x)^2$ and so on). As these masses
follow IH pattern we can express $m_1$ and $m_2$ in terms of $m_3$ and the mass square
differences $ \Delta m_{21}^2$ and  $ \Delta m_{31}^2$ as
\bea
m_1^2 &=& m_3^2 + \Delta m_{13}^2\nn\\
m_2^2 &=& m_3^2 + \Delta m_{21}^2 + \Delta m_{13}^2\;.
\eea
Now squaring both sides of  Eq. (\ref{eqm11a}) and substituting the values of
$m_1$, $m_1^2$ and $m_2^2$ we obtain
\bea
Pm_3^2 +Q m_3+R=0\;,
\eea
with
\bea
P=a^2+c^2-b^2,~~~~~~~Q=2 a c \sqrt{\Delta m_{13}^2},~~~~~~~~R=a^2 \Delta m_{13}^2 - b^2(\Delta m_{13}^2
+ \Delta m_{21}^2)\;.
\eea
This is a quadratic equation in $m_3^2$ which can be solved numerically. Now using
the allowed values of the mass square differences, in Figure-2, we show the variation of
$m_3$ with the perturbation parameters $x$ and $y$. From the figure it can be seen that
the mass scale which is $\sim {\cal O}(10^{-2})$, decreases with the increase of $x$ and $y$.

\begin{figure}[htb]
\includegraphics[width=8cm,height=6cm, clip]{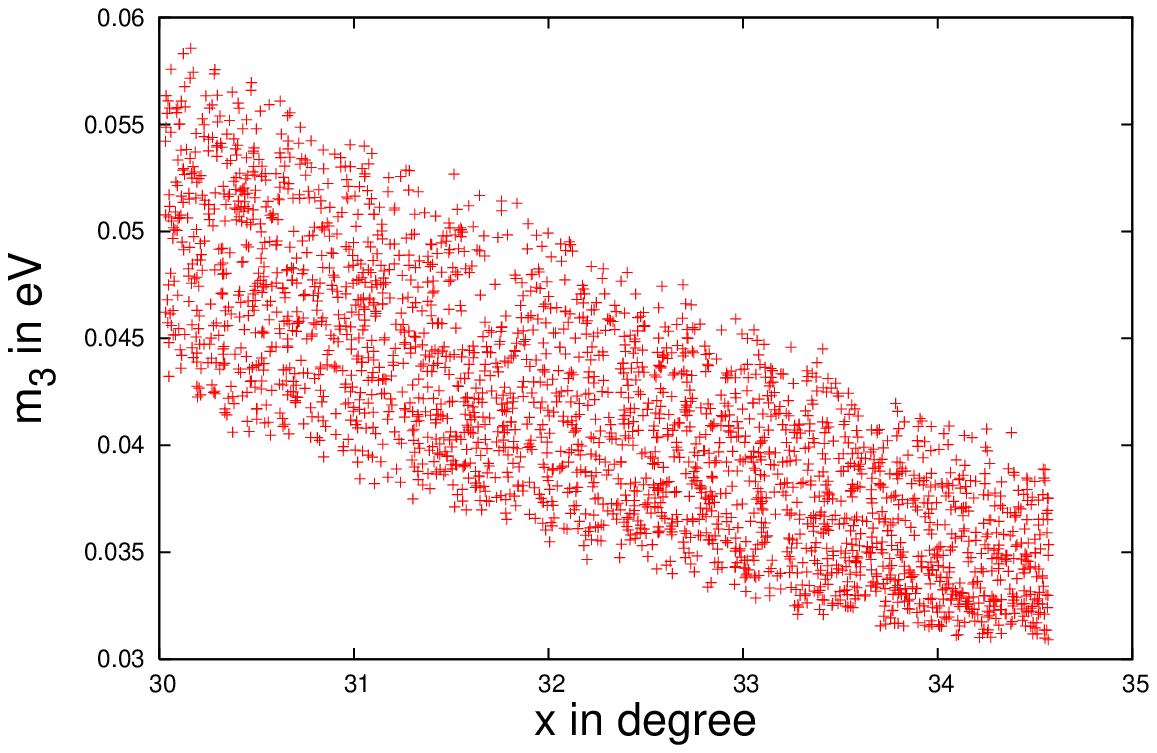}
\hspace{0.2 cm}
\includegraphics[width=8cm,height=6cm, clip]{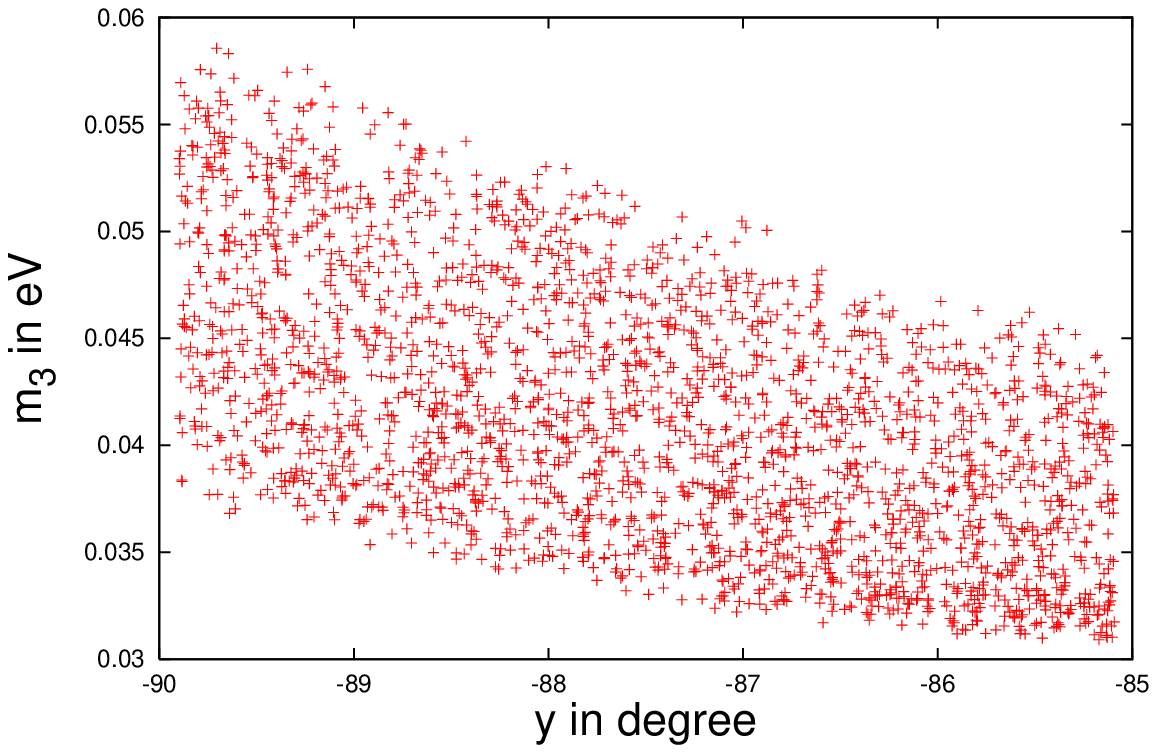}
\caption{Variation of the lightest neutrino neutrino mass ($m_3$ for IH)  with
the perturbation parameter $x$
(left panel) and $y$ (right panel).}
\end{figure}

{\bf{Case 2: $(M_{\nu})_{23}=0$}}

Now equating the (2,3) element of the neutrino mass matrix (\ref{mass}) to zero, we obtain the condition
\bea
&&m_{2}\sin^2x + \cos^2y (-3m_{3}+ 2m_{1}\sin^2x) +\sqrt{2}(m_{1}-m_{2})\cos y
\sin 2x \nn\\
&&+2m_{3}\sin^2y-3 m_{1}\sin^2x \sin^2y +\cos^2x(m_{1}+2m_{2}\cos^2y -3m_{2}\sin^2y)=0\;.
\eea
Since the above equation also involves the mass terms of all the three neutrinos, it is not possible
to infer any definite result regarding the ratio of neutrino masses. However proceeding as in the
previous case, one can obtain in the $m_{1}\rightarrow 0$ limiting case i.e., for
normal hierarchy
\be
\frac{m_2}{m_3}=\left |\frac{ 2 \cos 2 y + \cos^2 y}{\sin^2 x - \sqrt 2 \cos y \sin 2 x + \cos^2 x(2 \cos 2 y -
\sin^2 y)}\right |\;.
\ee

Similarly taking the limit $m_3 \to 0$ for Inverted Hierarchy case we obtain the following mass ratio
\be
\frac{m_1}{m_2}=\left |\frac{\sin^2 x - \sqrt{2} \cos y \sin 2 x +  \cos^2 x(2 \cos 2y-\sin^2 y)}
{\sin^2 x(\sin^2 y - 2 \cos 2 y) - \sqrt{2} \cos y \sin 2 x - \cos^2 x}\right |\;.
\ee
Now substituting the allowed ranges of $x$ and $y$ we obtain
\bea
&& \dfrac{m_{2}}{m_{3}}> 1 ~~~~~{(\rm for ~~NH)}\nn\\
&&\dfrac{m_{1}}{m_{2}}> 1~~~~~{(\rm for ~~IH)}\;.
\eea
Thus, the above results implied that there is no feasible solution or in other words
the neutrino masses do not satisfy either normal or inverted hierarchy pattern.

{\bf{Case 3: $(M_{v})_{22}=0$}}

Taking the $(M_\nu)_{22}$ element to be zero we obtain
\bea
&& m_{1}(\cos x + \sqrt 2 \sin x \cos y - \sqrt 3 \sin x \sin y)^2
+ m_3 \left ( \frac{\cos y}{\sqrt 2} + \frac{\sin y}{\sqrt 3} \right )^2\nn\\
&&+m_2( \sin x + \sqrt 3 \cos x \sin y - \sqrt 2 \cos x \cos y)^2=0\;.
\eea
Proceeding as in the previous cases we obtain
\bea
&& \dfrac{m_{2}}{m_{3}}> 1 ~~~~{\rm (for~~NH)}\nn\\
&& \dfrac{m_{1}}{m_{2}}< 1~~~~~{\rm (for~~IH)}\;.
\eea
Thus, the above analysis implies that the neutrino mass structure will follow Inverted mass hierarchy.

{\bf{Case 4: $(M_{\nu})_{33}=0$}}

Similarly equating the  $(M_\nu)_{33}$ element to zero we obtain
\bea
&& m_{1}(\cos x + \sqrt 2 \sin x \cos y + \sqrt 3 \sin x \sin y)^2
+ m_3  ( \sqrt 3 \cos y - \sqrt 2\sin y  )^2\nn\\
&&+m_2( \sin x - \sqrt 2 \cos x \cos y - \sqrt 3 \cos x \sin y)^2=0\;.\label{eqm33}
\eea
Now, substituting the allowed values of $x$ and $y$ we obtain
\be
\dfrac{m_{2}}{m_{3}}< 1
\ee for normal hierarchy and
\be \dfrac{m_{1}}{m_{2}}> 1,
\ee for
Inverted Hierarchy ($m_{3}\rightarrow 0$). Thus, implying that the neutrino mass matrix having texture one-zero with vanishing $(M_\nu)_{33}$ obeys Normal Hierarchy.

Now since
we know that texture-zero type with $(M_{\nu})_{33}$ supports normal mass hierarchy
pattern, it is possible to determine the absolute mass scale of the lightest neutrino
i.e., $m_1$ from Eq. (\ref{eqm33}). Proceeding as in the previous case and
substituting
 $m_2^2$ and $m_3^2$ as
\bea
m_2^2 &=& m_1^2 + \Delta m_{21}^2\nn\\
m_3^2 &=& m_1^2 + \Delta m_{31}^2 \;,
\eea
 we obtain the quadratic equation on $m_1$
\bea
P_1 m_1^2 +Q_1 m_1+R_1=0\;,
\eea
with
\bea
P_1=a^2+c^2-b^2,~~~~~~~Q_1=2 a c \sqrt{\Delta m_{31}^2},~~~~~~~~R_1=c^2 \Delta m_{31}^2 - b^2\Delta m_{21}^2
\;.
\eea
This is a quadratic equation in $m_1^2$ which can be solved numerically. Now using
the allowed values of the mass square differences, in Figure-3, we show the variation of
$m_1$ with the perturbation parameters $x$ and $y$. From the figure it can be seen that
the mass scale which is $\sim {\cal O}(10^{-2})$, increases with the increase of $y$ and it
is insensitive to the variation of $x$.

\begin{figure}[htb]
\includegraphics[width=8cm,height=6cm, clip]{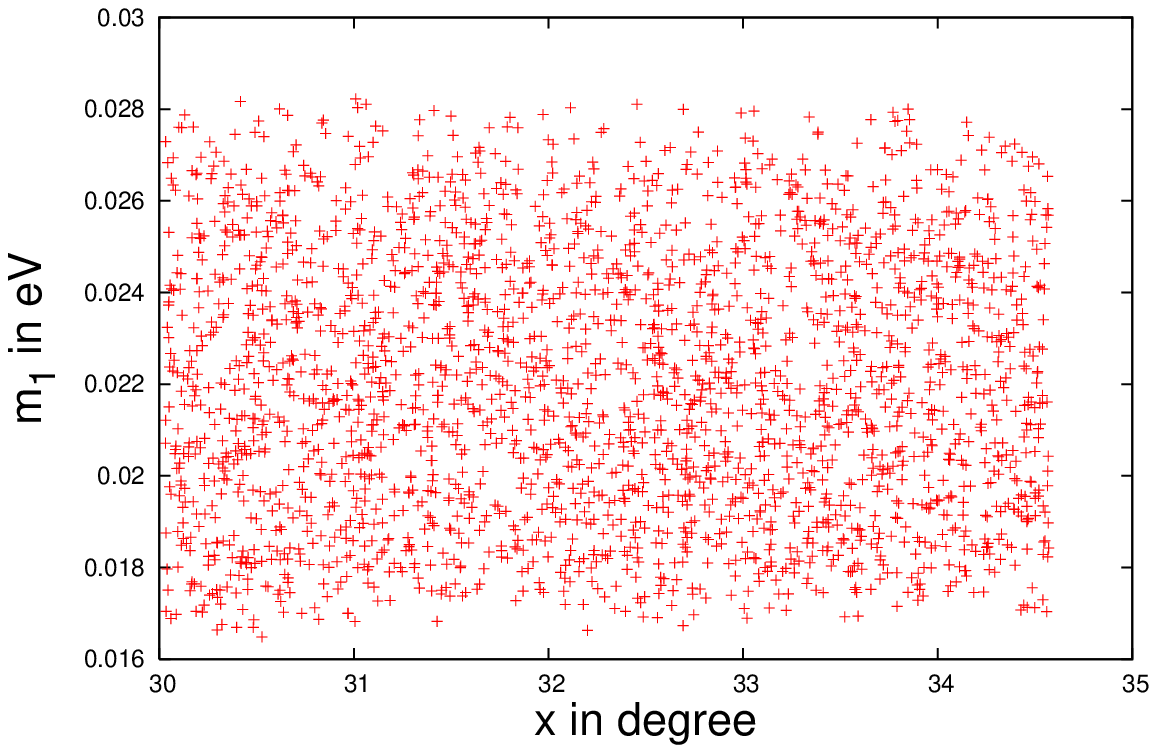}
\hspace{0.2 cm}
\includegraphics[width=8cm,height=6cm, clip]{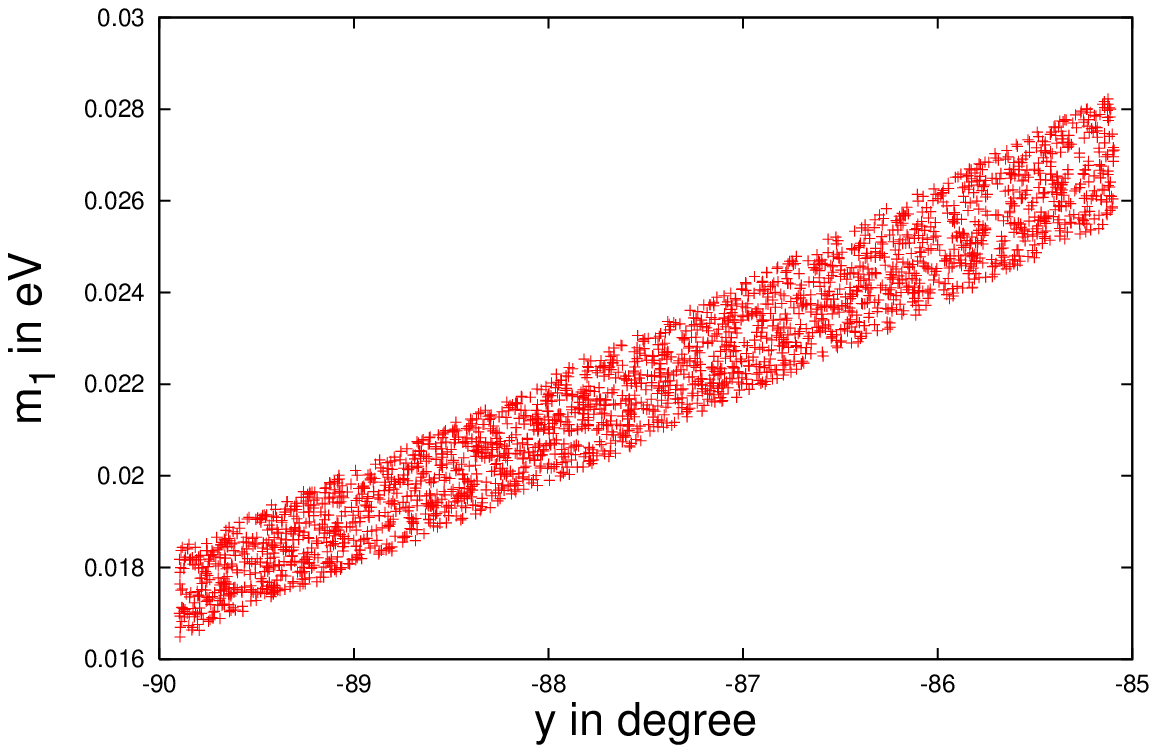}
\caption{Variation of the lightest neutrino mass scale ($m_1$)  with
the perturbation parameter $x$
(left panel) and $y$ (right panel).}
\end{figure}

{\bf{Case 5: $(M_{\nu})_{12}=0$} }

Similarly equating the  $(M_\nu)_{12}$ element to zero we obtain
\bea
&& m_{1}(  \cos y +\sin y-\sqrt 2 \cos x - )(\cos x + \sin x (\sqrt 2 \cos y - \sqrt 3 \sin  y))\nn\\
&&+m_2( \cos x \cos y + \sqrt 2 \sin x) (-\sin x + \sqrt 3 \cos x \cos y - \sqrt 2  \cos x\sin y )\nn\\
&&+ m_3 \sin y  ( \sqrt 3 \cos y + \sqrt 2\sin y  )=0\;.\eea
Now, substituting the allowed values of $x$ and $y$ we obtain
\be
\dfrac{m_{2}}{m_{3}}> 1
\ee for normal hierarchy and
\be \dfrac{m_{1}}{m_{2}}< 1,
\ee for
Inverted Hierarchy ($m_{3}\rightarrow 0$). Thus, implying that the neutrino mass matrix having texture one-zero with vanishing $(M_\nu)_{12}$ obeys Inverted Hierarchy.

{\bf{Case 6: $(M_{\nu})_{13}=0$}}

Similarly equating the  $(M_\nu)_{13}$ element to zero we obtain
\bea
&& m_{1}( \sqrt 2 \cos x -  \sin x \cos y)(-\cos x -\sin x(\sqrt 2 \cos y + \sqrt 3 \sin  y))\nn\\
&&+m_2( \cos x \cos y+ \sqrt 2 \sin x)(-\sin x + \cos x (\sqrt 2 \cos y+\sqrt 3  \sin y))\nn\\
&&+ m_3 \sin y  (- \sqrt 3 \cos y + \sqrt 2\sin y  )=0\;.
\eea
Now, substituting the allowed values of $x$ and $y$ we obtain
\be
\dfrac{m_{2}}{m_{3}}< 1
\ee for normal hierarchy and
\be \dfrac{m_{1}}{m_{2}}> 1,
\ee for
Inverted Hierarchy ($m_{3}\rightarrow 0$). Thus, implying that the neutrino mass matrix having texture one-zero with vanishing $(M_\nu)_{13}$ obeys Normal Hierarchy.

Similarly, the above analysis is repeated for all the perturbations and the results
 are presented in Table-1. The patterns for which there does not exist any feasible solution
 for any texture one-zero cases are not listed in the Table.

Next we will consider the perturbation of the form $V_{\alpha} \cdot V_{ij}$. It has already been shown
in Ref. \cite{rf10} that perturbation of this form with $\alpha$=BM and DC are excluded because
of the following reasons.\\
{\bf i.} $V_{BM}.V_{12}$ gives $\theta_{13}=0$ and hence excluded by the recent MINOS and T2K results.\\
{\bf ii.}  $V_{BM}.V_{23}$ gives $\theta_{12}> 45^\circ $ and hence excluded by solar neutrino data\\
{\bf iii.} $V_{BM}.V_{13}$ gives $\theta_{13}\in [26.7^\circ, 33.7^\circ]$ and hence excluded by the recent MINOS and T2K results.\\
{\bf iv.} $V_{DC}.V_{12}$ gives $\theta_{13}=0$ and hence excluded by the recent MINOS and T2K results.\\
{\bf  v.}  $V_{DC}.V_{23}$ gives $\theta_{13}\in [26^\circ, 35^\circ] $ and hence excluded by the recent MINOS and T2K results.\\
{\bf vi.} $V_{DC}.V_{13}$ gives $\theta_{13}>45^\circ $ and hence excluded.\\

For the perturbation of the type $ V_{TB}\cdot V_{ij}$, we found that\\
{\bf i.} For $ij=12$, $\sin \theta_{13}$
turns out to be zero and hence not allowed by the current data.\\
{\bf ii.} For $ij=23 $, we obtain the mixing angles by comparing both sides of the relation
$V_{PMNS}= V_{TB}\cdot V_{ij}$ as
\bea \tan \theta_{12}= \frac{\cos y}{\sqrt 2} \;,
\eea
which gives the value of the perturbation angle $y= \left (13.6_{-6.0}^{+7.0} \right )^\circ$.
We find that for such a perturbation the only allowed possibility is $(M_{\nu})_{33}=0$ texture-zero,
which supports normal hierarchy.\\
  {\bf iii.} For $ij=13$, we obtain  $z= \left (13.7_{-6.2}^{+7.3} \right )^\circ$, for which there
is no feasible solution that exists for any type of texture one-zero.

Similar analysis is performed for the perturbation of the form $V_{ij}\cdot V_{\alpha}$ and
it is found that  $\alpha$= BM and DC cases are excluded  as in the previous case due to the
following reasons.\\
{\bf i.} $V_{12} \cdot V_{BM}$ gives $x=-\left (14.69_{-1.3}^{+1.4} \right )^\circ $ and no feasible solution
is found for such perturbation.\\
{\bf ii.}  $V_{23} \cdot V_{BM}$ gives $\theta_{13}=0$ and hence excluded by the recent MINOS and T2K results.\\
{\bf iii.} $V_{13}\cdot V_{BM}$ gives $z=\left (14.69_{-1.33}^{+1.38} \right )^\circ $ and no feasible solution
is found for such perturbation.\\
{\bf iv.} $V_{12}\cdot V_{DC}$ gives $\theta_{13}\in [13.5^\circ, 45^\circ]$ and hence excluded by the recent MINOS and T2K results.\\
{\bf  v.}  $V_{23}\cdot V_{DC}$ gives $\theta_{13}=0 $ and hence excluded by the recent MINOS and T2K results.\\
{\bf vi.} $V_{13}\cdot V_{DC}$ gives $\theta_{23}>54.7^\circ $ and hence excluded by the atmospheric neutrino data.\\

Furthermore, for $\alpha$=TB among all possibilities, we observe that only $V_{23} \cdot V_{TB}$ has a
 feasible solution for  $(M_{\nu})_{11}=0$  and it supports Inverted hierarchy.

\begin{table}
\caption{The allowed mass matrix texture for different type of perturbations to BM/TBM/DC mixing matrices.}
\begin{small}
\begin{tabular}{|c|c|c|c|c|c|c|c|c|}
\hline
&Perturbation& Angles & Case-1 & Case-2 & Case-3 & Case-4 & Case-5 & Case-6\\
&Scenarios&(in degree)& ${\small  (M_{\nu})_{11}=0}$ & {\small $(M_{\nu})_{23}=0$}
&{\small $(M_{\nu})_{22}=0$} &{\small $(M_{\nu})_{33}=0$ }&
{\small $(M_{\nu})_{12}=0$}&{\small  $(M_{\nu})_{13}=0$}\\
\hline
1&$ V_{TB}.V_{23}.V_{12} $&$x=33.2^{+1.4}_{-3.2} $&IH&~No soln~&IH& NH & IH & NH\\
& &$y= -\left (88.2_{-3.1}^{+1.7} \right )$& & & & & &\\
\hline
2&$V_{TB}.V_{13}.V_{12}$&$x=- \left (53.7_{-2.1}^{+0.7} \right )$ & IH &NH & NH & IH & No Soln & IH\\
& &$z= 88.7_{-2.2}^{+1.2} $& & & & & & \\
\hline
3 &$V_{TB}.V_{23}.V_{13}$&$y=18.3_{-5.7}^{+4.4}$& ~ No soln~ & ~No soln~ & IH & NH & IH& NH\\
& &$z=25.1^{+2.0}_{-4.2}$& & & & & &\\
\hline
4&$V_{23}.V_{13}.V_{TB}$&$z=-1.1 \pm 1.4 $& ~No soln~ &
 ~No soln~ &~ No soln~&~No soln~ & NH & NH\\
& &$y = 87.8_{-3.8}^{+2.1} $& & & & & &\\
\hline
5&$V_{23}.V_{12}.V_{TB}$&$x=-1.1 \pm 1.4 $& ~No soln~ &~No soln~&~No soln~
&~No soln~ & NH & NH \\
& &$y= 87.8_{-3.8}^{+2.1} $& & & & & & \\
\hline
6&$V_{BM}.V_{13}.V_{12}$&$z=3.1_{-5.4}^{+5.3}$& NH & IH & IH & IH & NH & No Soln\\
& &$x= 45.2_{-0.1}^{+1.3}$& & & & & & \\
\hline
7&$V_{BM}.V_{23}.V_{12}$&$y=-3.1_{-5.3}^{+5.4}$ & NH & IH & IH & IH & IH & IH\\
& &$x= 44.8_{-1.3}^{+0.1}$& & & &  & & \\
\hline
8&$V_{23}.V_{12}.V_{BM}$&$x=-10.5\pm 1.0 $&No soln&No soln&No soln&IH &NH & No Soln\\
& &$y=-1.8_{-3.6}^{+3.8}$& & & &  & &\\
\hline
9&$V_{DC}.V_{13}.V_{12}$&$x=51.6_{-3.2}^{+3.5}$& NH & IH & IH & IH & NH & IH\\
& &$z=16.6_{-5.2}^{+5.1}$& & & & & & \\
\hline
10&$V_{DC}.V_{23}.V_{13}$&$y=37.9_{-1.9}^{+1.8}$& IH & IH & IH & NH & IH  & NH\\
& &$z=12.7_{-1.7}^{+1.1}$& & & & & & \\
\hline
11&$V_{23}.V_{13}.V_{DC}$&$y=-12.6 \pm 3.9$&No soln&No soln&No soln &
No soln &  No Soln & NH\\
& &$z=-12.8 \pm 1.2 $& & & & & &\\
\hline
12&$V_{23}.V_{12}.V_{DC}$&$x=-\left (17.8_{-1.6}^{+1.2}\right )$& NH & IH & IH & No soln & No Soln & NH\\
& &$y=-\left (10.6_{-3.5}^{+3.6}\right )$& & & & & & \\
\hline
13&$V_{13}.V_{12}.V_{DC}$&$x=34.96_{-6.17}^{+5.65}$& IH & NH & NH & No soln & No Soln  & IH \\
& &$z=-6.4 \pm 0.1$& & & & & & \\
\hline
\end{tabular}
\end{small}
\end{table}

\begin{table}
\caption{Continuation of Table-1.}
\begin{small}
\begin{tabular}{|c|c|c|c|c|c|c|c|c|}
\hline
&Perturbation& Angles & Case-1 & Case-2 & Case-3 & Case-4 & Case-5 & Case-6\\
&Scenarios&(in degree) & & & & & &\\
\hline
14&$V_{TB}.V_{23}$ & $ y=13.6_{-6.0}^{+7.0}$&~ No Soln~ &~ No Soln~ & ~No Soln~ & NH &~ No Soln~ & NH  \\
\hline
15&$V_{23}.V_{TB}$ & $ y=13.6_{-6.0}^{+7.0}$& IH & No Soln & No Soln & ~No Soln~ & ~No Soln~ &~ No Soln ~\\
\hline
16&$V_{TB}.V_{13}$ & $ z=13.7_{-6.2}^{+7.3}$& No Soln & No Soln & No Soln & No Soln & No Soln & NH \\
\hline
17&$V_{12}.V_{BM}$ & $ x=- \left (14.7_{-1.3}^{+1.4}\right )$& No Soln & No Soln & No Soln & No Soln & NH & No Soln \\
\hline
18&$V_{13}.V_{BM}$ & $ z=14.7_{-1.4}^{+1.3}$& No Soln & No Soln & No Soln & No Soln & No Soln & NH \\
\hline
\end{tabular}
\end{small}
\end{table}

To summarize, motivated by the recent data from T2K and MINOS which show the evidence of a relatively
large $\theta_{13}$  at $3 \sigma$ level, we study the phenomenological implications of
BM/TBM/DC mixing matrices with some  modifications. It has been shown in Ref. \cite{rf10}
that such modified mixing matrices could accommodate the  observed large $\theta_{13}$.
Using such mixing matrices and assuming texture one-zero in the neutrino mass matrices
we have shown that whether such mixing matrices satisfy normal or inverted mass hierarchy, or
phenomenologically viable or not.

{\bf Acknowledgments}
KND  would like to thank University Grants Commission for financial support.
The work of RM was partly supported by the Council of Scientific and Industrial Research,
Government of India through grant No. 03(1190)/11/EMR-II.

\end{document}